\documentclass{article}

\usepackage{arxiv}

\usepackage[utf8]{inputenc} 
\usepackage[T1]{fontenc}    
\usepackage{hyperref}       
\usepackage{url}            
\usepackage{booktabs}       
\usepackage{amsfonts}       
\usepackage{nicefrac}       
\usepackage{microtype}      
\usepackage{lipsum}		
\usepackage{graphicx}
\usepackage{amsmath}
\usepackage{amssymb}
\usepackage{mathptmx}
\usepackage{enumitem}
\usepackage{tikz}
\usepackage{empheq}
\usepackage{graphics}
\usepackage{color}
\usepackage{subfigure}
\usepackage{tikz}
\usepackage{graphicx}
\usepackage{dcolumn}
\usepackage{bm}

\newtheorem{theorem}{Theorem}
\newtheorem{definition}[theorem]{Definition}

\title{On the Stochastic Flows on\\ $(m+n+1)$-Dimensional Exotic Spheres}

\author{Nurfa Risha\\
  Departemen Fisika, Fakultas Matematika dan Ilmu Pengetahuan Alam\\
  Universitas Gadjah Mada\\
  Sekip Utara Bulaksumur, Yogyakarta, 55281, Indonesia\\
  \\
  Program Studi Pendidikan Fisika, Fakultas Matematika dan Ilmu Pengetahuan Alam\\
  Universitas Pendidikan Ganesha\\Jalan Udayana No. 10-11, Singaraja 81116, Indonesia.\\
  \texttt{nurfa.risha@undiksha.ac.id}\\
   \And
   Adhitya Ronnie Effendie\\
 Departemen Matematika, Fakultas Matematika dan Ilmu Pengetahuan Alam\\
 Universitas Gadjah Mada\\
 Sekip Utara Bulaksumur, Yogyakarta, 55281, Indonesia\\
 \texttt{adhityaronnie@ugm.ac.id}\\
  \AND
  Muhammad Farchani Rosyid\\
  Departemen Fisika, Fakultas Matematika dan Ilmu Pengetahuan Alam\\
  Universitas Gadjah Mada\\
  Sekip Utara Bulaksumur, Yogyakarta, 55281, Indonesia\\
  \texttt{farchani@ugm.ac.id}\\
}

\begin{document}
\maketitle

\begin{abstract}
Stochastic flows of Stratonovich stochastic differential equations on exotic spheres have been studied. The consequences of the choice of exotic differential structure on stochastic processes taking place on the topological space $S^{m+n+1}$ as state space of the processes have been investigated. More precisely, we have investigated the properties of stochastic processes where the state spaces of the stochastic processes under consideration are $({m+n+1})$-dimensional differentiable manifolds which are homeomorphic but not necessarily diffeomorphic to standard ${(m+n+1)}$-dimensional sphere. The differentiable manifolds have been constructed from disjoint union $\mathbb{R}^{m+1}\times S^{n}\sqcup S^m\times \mathbb{R}^{n+1}$ by identifying every pair of its points using a map $u :\mathbb{R}^{m+1}\times S^n\rightarrow S^m\times \mathbb{R}^{n+1}$ which is constructed from a diffeomorphism $h_1\times h_2:S^m\times S^n\rightarrow S^m\times S^n$. The diffeomorphisms $h_1$ and $h_2$, therefore, can be regarded as the carriers of the "exoticism" of the constructed manifolds. For all of the above purposes, homeomorphisms $h$ from the above-constructed manifolds onto the standard sphere explicitly in term of the diffeomorphisms $h_1$ and $h_2$ have been constructed. Using the homeomorphisms $h$ and all their associated maps derived from them and expressed in terms of $h_1$ and $h_2$ as well as their derivatives, we construct the stochastic processes or flows on the above-constructed manifolds corresponding to stochastics processes on the standard sphere $S^{m+n+1}_s$. The stochastic processes yielded from the above construction on the constructed manifolds can be regarded as the same stochastic processes on $S^{m+n+1}_s$ but described in exotic differential structures on $S^{m+n+1}$.
\end{abstract}


\section{Introduction}
$\qquad$Stochastic processes appear in many physical phenomena, such as the well-known diffusion process, heat conduction, Brownian motion, the formation of river meander, the fluctuation current due to thermal noise, etc. Mathematically, a stochastic process is a set of random variables parametrized by another set (discrete or continuous) playing the role of time. A stochastic process is a solution of so-called stochastic differential equations. Since there exists a distribution function for every random variable, there is an associated distribution function depending on the time for every stochastic process. The distribution function associated with a stochastic process satisfies a deterministic partial differential equation called the Fokker-Planck equation associated with the process which is obtained from the stochastic differential equation governing the process via It\^{o} equation. A Fokker-Planck equation is used by Einstein to define the time evolution of the probability density of particles whereas Langevin rewrites an explicit deterministic momentum equation for a particle augmented by a Gaussian white noise force which perturbs the particle trajectory. The approach of Langevin now constitutes a canonical stochastic differential equation for a full scope of systems, but since its introduction, the mathematical and physical interpretation of the white noise term has been discussed and debated. The white noise interpretation of stochastic differential equations naturally leads to stochastic differential equations in Stratonovich sense. This is because discrete-time and smooth approximations of white noise drive stochastic differential equations to converge to stochastic differential equations in Stratonovich sense, not in It\^{o} sense.

$\qquad$The microscopic description of the dynamics of a diffusion process on a connected compact differentiable manifold $M$, for instance, is typically represented by a so-called Stratonovich formulation of stochastic differential equations \cite{book Ikeda and watanabe}. Likewise, the dynamic descriptions of the other stochastic processes on differentiable manifolds are also represented in the formulation. The advantage of this formulation is that It\^{o}'s formula appears in the same form as the fundamental theorem of calculus; therefore, stochastic calculus in this formulation takes a more familiar form. The other advantage of Stratonovich formulation, which is more important and relevant to our investigation is that stochastic differential equations on manifolds in this formulation transform consistently under diffeomorphisms between manifolds \cite{book Hsu}. Here we will consider a Stratonovich's stochastic differential equation on compact Riemannian manifold. There are at least two ways to solve the differential equation. Firstly, we solve it locally (i.e. all mathematical terms appearing in the equation are represented locally concerning coordinate systems covering the manifold, and these local representations of the equation are solved simultaneously) and then patch them up. In the second one, we obtain the solution of the equation in a suitable embedding space $\mathbb{R}^N$ and prove that the solution stays in $M$ whenever the starting point $q$ is a point in $M$. Both methods work well only in the case that the vector fields appearing in the above differential equation as coefficients are smooth (at least $C^2$-differentiable). In the case of non smooth drifting vector field, i.e. it has only Sobolev regularity and bounded divergence, Zhang \cite{Zhang} has point out the existence and uniqueness of a so-called invertible $\nu $-almost everywhere stochastic invertible flows of the stochastic differential equation, where $\nu $ is the Riemannian measure on the manifold under study. The starting point of Zhang's proof is the constructions of a sequence of smooth vector fields, which is convergent in Sobolev norm to the drifting vector field. The construction of such a sequence is possible because the drifting vector field is assumed to be a Sobolev vector field. The sequence of smooth vector fields leads to a sequence of stochastic differential equations which "converges" to the initial stochastic differential equation. Every stochastic differential equations in the sequence have a unique solution. Assuming the boundedness of the drifting vector field and its divergence, the sequence of the solutions of the stochastic differential equations in the sequence converges in a suitable metric constructed from geodesic metric to the unique invertible $\nu $-almost everywhere stochastic invertible flows.

$\qquad$In general, the laws of physics must be independent of the choice of a coordinate system. This statement means, for instance, that there must be a family of coordinate systems that are compatible with describing space-time, i.e. the space of all events. A set of compatible coordinate systems in space-time is a subset of a so-called differential structure. Space-time is an example of a topological space equipped with a differential structure \cite{paper Freedman}. In a topological space, we may find more than one differential structures. Therefore, from a single topological space of all events, we can construct more than one space-time with different differential structures. If the yielded space-times are not distinguishable in the sense that they are not diffeomorphic, they may lead to inequivalent formulations of the law of physics. The totality of inequivalent differential structures on a topological space is called the {\em exotica of differential structures} on the topological space, and a differential structure that is not equivalent to the standard one is called an exotic differential structure on the topological space under consideration \cite{Torsten,Brans}. Unfortunately, among the exotic differential structures whose existences are already known on various topological spaces, only the exotic differential structures on spheres can be constructed explicitly. Therefore, the exotic spheres can be used as toy models serving as configuration or phase spaces of physical systems as well as the base manifolds for various field theories. Various problems of real interest can be answered by studying the models and the results they are compared to those obtained from the standard spheres \cite{Brans}.

$\qquad$In mathematics, spheres are topological spaces which are interesting to investigate. They serve, for instance, as models for configuration spaces of some mechanical systems. In physics, for example, the standard $7$-dimensional sphere $S^7_s$ is particularly interesting in related to supersymmetry breaking \cite{Englert} and to the work of Witten \cite{Edward Witten} in which he used it to cancel the global gravitational anomalies in $1985$. The standard seven-dimensional sphere is also regarded as the total space of a principal $SU(2)$ bundle of Yang-Mills theory \cite{Kengo1,Kengo2}. Two differential structures on  $7$-dimensional sphere $S^7$ are said to be equivalent if there is a diffeomorphism pulling back the differentiable maximal atlas from the second to the first. The connected compact topological space $S^7$ has more than one distinct differential structure that are not equivalent each other in this sense or more precisely there are connected compact $7$-dimensional differentiable manifolds which are homeomorphic but not diffeomorphic to the standard seven-sphere $S^7_s$ \cite{paper Milnor,paper Milnor1}. The differential structure on the standard sphere $S^7_s$ is called a standard differential structure, while the differential structures that are not equivalent to the standard one are called exotic differential structures. The topological space $S^7$ equipped with an exotic differential structure or a seven-dimensional differentiable manifold which is homeomorphic but not diffeomorphic to standard seven-dimensional sphere $S^7_s$ is called exotic $7$-sphere.

$\qquad$The main purpose of this work is to understand the consequences of the choice of exotic differential structure on stochastic processes taking place on the topological space $S^{m+n+1}$ as state space of the prosesses. More precisely, we investigate the properties of stochastic processes where the state spaces of the stochastic processes under consideration are $({m+n+1})$-dimensional differentiable manifolds $M^{m+n+1}_{(h_1,h_2)}$ which are homeomorphic but not necessarily diffeomorphic to standard ${(m+n+1)}$-dimensional sphere $S^{m+n+1}_s$. A manifolds $M^{m+n+1}_{(h_1,h_2)}$ is constructed (see \cite{paper Milnor}) from disjoint union $\mathbb{R}^{m+1}\times S^{n}\sqcup S^m\times \mathbb{R}^{n+1}$ by identifying every pair of its points using a map $u :\mathbb{R}^{m+1}\times S^n\rightarrow S^m\times \mathbb{R}^{n+1}$ which is constructed from a diffeomorphism $h_1\times h_2:S^m\times S^n\rightarrow S^m\times S^n$. Whether a manifold $M^{m+n+1}_{(h_1,h_2)}$ is homeomorphic or even diffeomorphic to the standard sphere $S^{m+n+1}_s$ depends on the diffeomorphism $h_1, :S^m\rightarrow S^m$ and  $h_2:S^n\rightarrow S^n$. We consider here only the cases where the diffeomorphisms $h_1$ and $h_2$ lead to the manifolds $M^{m+n+1}_{(h_1,h_2)}$ which are at least homeomorphic to standard sphere. The diffeomorphisms $h_1$ and $h_2$ therefore can be regarded as the carriers of the "exoticism" of $M^{m+n+1}_{(h_1,h_2)}$. For all of the above purposes, we firstly construct from the diffeomorphisms $h_1$ and $h_2$ a homeomorphism $h:M^{m+n+1}_{(h_1,h_2)}\rightarrow S^{m+n+1}_s$ explicitly in term of the diffeomorphisms $h_1$ and $h_2$. The homeomorphism $h$ topologically identifies the topological space $M^{m+n+1}_{(h_1,h_2)}$ with $S^{m+n+1}$. For instance, if the diffeomorphisms $h_1$ and $h_2$ determining $M^{m+n+1}_{(h_1,h_2)}$ are the identity maps, then the corresponding homeomorphism $h$ is a diffeomorphism. Assuming a stochastic differential equation which satisfies certain regularities on the standard spheres, then associated to the equation there exists a unique stochastic flow on the standard sphere in the sense of Zhang \cite{Zhang}. Using the identification $h$, a stochastic flow can be constructed on $M^{m+n+1}_{(h_1,h_2)}$. The stochastic flow yielded from the above construction on $M^{m+n+1}_{(h_1,h_2)}$ can be regarded as the same stochastic flow on $S^{m+n+1}_s$ but described in exotic differential structures on $S^{m+n+1}$. In turn, by making use of the homeomorphism $h$ and associated maps derived from it (expressed in terms of $h_1$ and $h_2$ as well as their derivatives), we construct a stochastic differential equation on $M^{m+n+1}_{(h_1,h_2)}$ corresponding to the stochastics differential equation on the standard sphere $S^{m+n+1}_s$.  We investigate then the stochastic flow yielded in the above mentioned construction on the exotic spheres. The main result of our investigation is depicted than in Theorem \ref{theorem2}.

\section{Milnor's Construction of Exotic Spheres and their Homeomorphism into Standard Sphere}

$\qquad$Let $h_1\times h_2:S^m\times S^n\rightarrow S^m\times S^n$ be a diffeomorphisme from $S^m\times S^n$ onto itself so that every $(x,y)\in S^m\times S^n$ is maped into $(h_1(x), h_2(y))\in S^m\times S^n$. Furthermore, define a map $u :\mathbb{R}^{m+1}\times S^n\rightarrow S^m\times \mathbb{R}^{n+1}$ defined by $u(tx,y)=(h_1(x),t^{-1}h_2(y))$ for every $t\in (0,\infty )$ and $(x,y)\in S^m\times S^n$. The manifold $M^{m+n+1}_{(h_1,h_2)}$ is obtained from $\mathbb{R}^{m+1}\times S^{n}\sqcup S^m\times \mathbb{R}^{n+1}$ by gluing or matching every $(tx,y)\in \mathbb{R}^{m+1}\times S^{n}$ with its image $u(tx,y)=(h_1(x),t^{-1}h_2(y))$ under the map $u$. It is an $(m+n+1)$-dimensional manifold. Wether the obtained manifold $M^{m+n+1}_{(h_1,h_2)}$ is homeomorphic or even diffeomorphic to $S^{m+n+1}_s$ depends merely on the maps $h_1$ and $h_2$.

$\qquad$Now consider two maps defined by
\begin{eqnarray}\label{homeo}
\mathbb{R}^{m+1}\times S^n\ni (\tilde{x},\tilde{y})\mapsto\left(\frac{\tilde{x}}{\sqrt{1+\|\tilde{x}\|^2}}, \frac{\tilde{y}}{\sqrt{1+\|\tilde{x}\|^2}}\right)\in S^{m+n+1},
\end{eqnarray}
and
\begin{eqnarray}\label{homeo1}
S^m\times\mathbb{R}^{n+1}\ni (\bar{x},\bar{y})\mapsto\left(\frac{h_1^{-1}(\bar{x})}{\sqrt{1+\|\bar{y}\|^2}}, \frac{\|\bar{y}\|h_2^{-1}(\frac{\bar{y}}{\|\bar{y}\|})}{\sqrt{1+\|\bar{y}\|^2}}\right)\in S^{m+n+1}.
\end{eqnarray}
If 
\begin{equation}\bar{x}=h_1(\frac{\tilde{x}}{\|\tilde{x}\|})\hspace{.6cm}\text{   and}\hspace{.6cm}\bar{y}=\frac{h_2(\tilde{y})}{\|\tilde{x}\|},
\end{equation}
i.e. $(\tilde{x},\tilde{y})$ and $(\bar{x},\bar{y})$ represent the same point in $M^{m+n+1}_{(h_1,h_2)}$, then it is clear that $\|\bar{y}\|=\|\tilde{x}\|^{-1}$ and
\begin{eqnarray}
\left(\frac{\tilde{x}}{\sqrt{1+\|\tilde{x}\|^2}}, \frac{\tilde{y}}{\sqrt{1+\|\tilde{x}\|^2}}\right)=
\left(\frac{h_1^{-1}(\bar{x})}{\sqrt{1+\|\bar{y}\|^2}}, \frac{\|\bar{y}\|h_2^{-1}(\frac{\bar{y}}{\|\bar{y}\|})}{\sqrt{1+\|\bar{y}\|^2}}\right).
\end{eqnarray}
Therefore the maps $h:M^{m+n+1}_{(h_1,h_2)}\rightarrow S^{m+n+1}_s$ defined by Eq.\eqref{homeo} and Eq. \eqref{homeo1} is well defined.

$\qquad$Now let $(\tilde{x},\tilde{y})$ and $(\breve{x},\breve{y})$ be any two elements of $\in \mathbb{R}^{m+1}\times S^n$ with $(\tilde{x},\tilde{y}) \neq (\breve{x},\breve{y})$. If 

\begin{eqnarray}\label{injektive}
\left(\frac{\tilde{x}}{\sqrt{1+\|\tilde{x}\|^2}}, \frac{\tilde{y}}{\sqrt{1+\|\tilde{x}\|^2}}\right) = \left(\frac{\breve{x}}{\sqrt{1+\|\breve{x}\|^2}}, \frac{\tilde{y}}{\sqrt{1+\|\breve{x}\|^2}}\right),
\end{eqnarray}
then 
\begin{eqnarray}\label{injektive1}
\frac{\tilde{x}}{\sqrt{1+\|\tilde{x}\|^2}} =  \frac{\tilde{x}}{\sqrt{1+\|\breve{x}\|^2}}\hspace{.6cm}\text{and}\hspace{.6cm}\frac{\tilde{y}}{\sqrt{1+\|\tilde{x}\|^2}} =  \frac{\tilde{y}}{\sqrt{1+\|\breve{x}\|^2}}.
\end{eqnarray}
Since $\tilde{y}$ and $\breve{y}$ are elements of $S^n$, we have
\begin{eqnarray}\label{injektive2}
\sqrt{1+\|\tilde{x}\|^2} =  \sqrt{1+\|\breve{x}\|^2}.
\end{eqnarray}
From Eq.\eqref{injektive1} and Eq.\eqref{injektive2} we have $(\tilde{x},\tilde{y})= (\breve{x},\breve{y})$. It contradicts the above asummption. Therefore, the map $h$ is an injection.

$\qquad$Let $(\gamma, \kappa)\in S^{m+n+1}_s$. Furthermore let $(x,y)\in S^m\times S^n$ and $t\in [0,\infty )$ such that
\begin{eqnarray}\label{invers}
\gamma=\frac{tx}{\sqrt{1+t^2}}\hspace{.6cm}\text{and}\hspace{.6cm}\kappa =\frac{y}{\sqrt{1+t^2}}.
\end{eqnarray}
From Eq. \eqref{invers} we have

\begin{eqnarray}\label{te1}
\|\gamma\|^2(1+t^2)=t^2\|x\|^2=t^2,
\end{eqnarray}
and
\begin{eqnarray}\label{te2}
\|\kappa\|^2(1+t^2)=\|y\|^2=1,
\end{eqnarray}
since $x\in S^m$ and $y\in S^n$. From Eq.\eqref{te1} and Eq.\eqref{te2} we obtain then
\begin{eqnarray}
t=\sqrt{\frac{\|\gamma\|^2}{\|\kappa\|^2}}=\frac{\|\gamma\|}{\|\kappa\|},
\end{eqnarray}
\begin{eqnarray}
tx&=&\frac{\|\gamma\|}{\|\kappa\|}\;\;\gamma\sqrt{\frac{\|\kappa\|^2}{\|\gamma\|^2}+1}=\gamma\sqrt{1+\frac{\|\gamma\|^2}{\|\kappa\|^2}},
\end{eqnarray}
and
\begin{eqnarray}
y&=&\kappa\sqrt{1+\frac{\|\gamma\|^2}{\|\kappa\|^2}}.
\end{eqnarray}
This means the map $h$ is surjective and therefore it is a bijection.
Let $f$ denote the invers of $h$. Thus, $f:S^{m+n+1}\rightarrow M^{m+n+1}_{(h_1,h_2)}$ is given by
\begin{eqnarray}\label{definisif}
S^{m+n+1}\ni (\gamma, \kappa)&\mapsto &\begin{cases}
\left(\frac{\|\gamma\|}{\|\kappa\|}\;\;\gamma\sqrt{\frac{\|\kappa\|^2}{\|\gamma\|^2}+1}\;,\; \kappa\sqrt{1+\frac{\|\gamma\|^2}{\|\kappa\|^2}} \right)\in \mathbb{R}^{m+1}\times S^n
\\
\\
\left(h_1\left[\gamma\sqrt{\frac{\|\kappa\|^2}{\|\gamma\|^2}+1}\;\right], \frac{\|\kappa\|}{\|\gamma\|}\;h_2\left[\kappa\sqrt{1+\frac{\|\gamma\|^2}{\|\kappa\|^2}}\;\right]\right)\in S^m\times \mathbb{R}^{n+1}.
\end{cases}
\end{eqnarray}

\section{Stochastic Flows on $M^{m+n+1}_{(h_1,h_2)}$}
$\qquad$Let $(W_t)_{t\geq 0}$ denote the $d$-dimensional standard Brownian motion on the classical Wiener space $(\Omega , \mathcal{F},P; (\mathcal{F}_t)_{t\geq 0})$, where $\Omega $ is the space of all continuous function from $\mathbb{R}_+$ to $\mathbb{R}^d$ with locally uniform convergence topology, $\mathcal{F}$ is the Borel $\sigma $-field on $\Omega $ generated by the topology, $P$ is the Wiener measure on $\mathcal{F}$, and $(\mathcal{F}_t)_{t\geq 0}$ is the filtration generated by $W_t(\omega )=\omega _t$. Now consider the Stratonovich's stochastic differential equation 
\begin{equation}\label{SDE1}
dq_t=X_0(q_t)dt+\sum _{k=1}^dX_k(q_t)\circ dW^k_t,\hspace{0.6cm}q_0=q
\end{equation}
on a compact differentiable Riemannian manifold $(M, g)$, where $X_i$, $i=0,... d$ are $d+1$ vector fields on $M$. There are at least two ways to solve the differential equation. In the first way, we solve it locally and then patch them up. In the second one, we obtain the solution of the equation in a suitable embeding space $\mathbb{R}^N$ and prove that the solution stays in $M$ whenever the starting point $q\in M$. It is well-known that the two ways as mentioned above work well only if the drift vector field $X_0$ and the other vectors $X_k$ $(k=1,...d)$ appearing in the above differential equation are smooth (at least $C^2$). For the case of $X_0$ having only Sobolev regularity and bounded divergence and $X_k$ $(k=1,\cdots,d)$ are $C^2$, Zhang \cite{Zhang} has shown the existence and uniqueness of the so-called $\nu $-almost everywhere stochastic invertible flows of SDE\eqref{SDE1}, where $\nu $ is the Riemannian measure on $M$.

\begin{definition}\label{DefZhang}\textbf{(Zhang \cite{Zhang})}
	Let $q_t(\omega ,q)$ be an $M$-valued measurable stochastic field on $\mathbb{R}_+\times \Omega \times M$. The flow $q_t(q)$ is called a $\nu $-almost everywhere stochastic flow of \eqref{SDE1} corresponding to vector fields $X_i$ ($i=0,...d$) if
	\begin{enumerate}
		\item For $\nu $-almost all $q\in M$, $t\mapsto q_t(q)$ is continuous and $(\mathcal{F}_t)$-adapted stochastic process and, satisfies that for any $t>0$ and $\zeta \in C^\infty (M)$
		\begin{equation}
		\zeta(q_t(q))=\zeta (q)+\int _0^tX_0\zeta (q_s(q))ds+\int _0^tX_k\zeta (q_s(q))\circ dW^k_s\hspace{.3cm}\forall t\geq 0.
		\end{equation}
		\item For arbritary $t\geq 0$ and $P$-almost all $\omega \in \Omega $, $(\nu \circ q_t)(\omega ,\cdot)\ll \nu $. For any $T>0$. there is a positive constant $K_{T,X_0,X_k}$ such that for all non-negative measurable function $\zeta $ on $M$
		\begin{equation}
		\sup _{t\in [0,T]}\mathbb{E}\int _M\zeta (q_t(q))\nu (dq) \leq K_{T,X_0,X_k}\int _M \zeta (q)\nu (dq).
		\end{equation}
	\end{enumerate}
	Furthermore, the $\nu $-almost everywhere stochastic flow $q_t(q)$ of \eqref{SDE1} is said to be invertible if for all $t\geq 0$ and $P$-almost all $\omega \in \Omega $, there exists a measurable invers $q_t^{-1}(\omega , \cdot )$ of  $q_t(\omega ,\cdot )$ so that $\nu \circ q_t^{-1}(\omega , \cdot )=\rho _t(\omega ,\cdot )\nu $, where the density $\rho _t(q)$ is given by
	\begin{equation}
	\rho _t(q)=\exp \left[\int _0^t \text{div} X_0(q_s(q))ds + \int _0^t \text{div} X_k(q_s(q))\circ dW^k_s \right].
	\end{equation}
\end{definition}

$\qquad$Let $C^k(TM)$ be the set of all $C^k$-differentiable vector fields on $M$, for every $k\in \mathbb{N}\cup \{\infty \}$. For every $p\geq 1$, we define
\begin{equation}
\|X\|_p:=\left[\int _M|X(q)|^p\nu (dq)\right]^{1/p},
\end{equation}
and
\begin{equation}
\|X\|_{1,p}:= \|X\|_p+\left[\int _M|\nabla X(q)|^p\nu (dq)\right]^{1/p},
\end{equation}
for every $X\in C^\infty (TM)$, where $\nabla $ is the Levi-Civita connection associated to the metric tensor $g$ on $M$. The completions of $C^\infty (TM)$ with respect to $\|\cdot \|_p$ and $\|\cdot \|_{1,p}$ will be denoted by $L^p(TM)$ and $\mathbb{H}^p_1(TM)$ respectively. Then Zhang found the following fact.
\begin{theorem}{\em \textbf{(Zhang \cite{Zhang})}}\label{TheZhang}
	Let $X_0\in L^\infty (TM)\cap \mathbb{H}^p_1(TM)$ for some $p\geq 1$, $\text{div } X_0\in L^\infty (M)$, and for each $k=1,\cdots ,d$, $X_k\in C^2(TM)$. Then there exists a unique $\nu$-almost everywhere stochastic invertible flow $q_t(q)$ of \eqref{SDE1} in the sense of Definition \ref{DefZhang}
\end{theorem}

$\qquad$Now consider the case where $M$ is the standard sphere $S^{m+n+1}_s$ provided by the natural Riemannian metric tensor $g$ induced from Euclidean metric in $\mathbb{R}^{m+n+2}$. Let $V_0\in L^\infty (TS^{m+n+1}_s)\cap \mathbb{H}^p_1(TS^{m+n+1}_1)$ and $V_k$ ($k=1,\cdots , d$) be $C^2$-vector fields on $S^{m+n+1}$. Then according to Theorem \ref{TheZhang}, there exists a unique $\nu$-almost everywhere stochastic invertible flow $q_t(q)$ of the following stochastic differential equation
\begin{equation}\label{SDE2}
dq_t=V_0(q_t)dt+\sum _{k=1}^dV_k(q_t)\circ dW^k_t,\hspace{0.6cm}q_0=q
\end{equation}
on $S^{m+n+1}_s$. With the identivication $h$ and its invers $f$ constructed in the previous section, then $(f\circ q_t\circ h)(\bar{q})$ is a stochastic flow on $M^{m+n+1}_{h_1,h_2}$, where $\bar{q}$ is the element of $M^{m+n+1}_{h_1,h_2}$ which is identified with $q$, i.e., $h(\bar{q})=q$. The stochastic flow $(f\circ q_t\circ h)(\bar{q})$ and the stochastic flow $q_t(q)$ is actually the same flow on the topological space $S^{m+n+1}$. The flow $q_t(q)$ is the stochastic flow on $S^{m+n+1}$ described using the standar differential structure on $S^{m+n+1}$. Whereas, the flow $(f\circ q_t\circ h)(\bar{q})$ can be regarded as the same stochastic flow on $S^{m+n+1}$ described using another differential structure which may be exotic (depending on the maps $h_1$ and $h_2$). If the maps $h_1$ and $h_2$ leads to manifold $M^{m+n+1}_{h_1,h_2}$ which is diffeomorphic to $S^{m+n+1}_s$ so that $h$ is a diffeomorphism, then $(f\circ q_t\circ h)(\bar{q})$ and $q_t(q)$ clearly are the same $\nu$-almost everywhere stochastic invertible flow on $S^{m+n+1}_s$. Now we will investigate the properties of the flow $(f\circ q_t\circ h)(\bar{q})$ for the general case in which $M^{m+n+1}_{h_1,h_2}$ may not be diffeomorphic but homeomorphic to $S^{m+n+1}_s$.

$\qquad$The vector fields $V_0,V_1, \cdots , V_d$ on $S^{m+n+1}_s$ appearing in equation \eqref{SDE2} are vector fields on $S^{m+n+1}$ when they are described using the standard differential structure on $S^{m+n+1}$. The vector fields are {\em seen as} $f_\ast V_0$, $f_\ast V_1$, $\cdots $, $f_\ast V_d$ on $M^{m+n+1}_{h_1,h_2}$ if they are described by using another (exotic) differential structure on $S^{m+n+1}$. The stochastic differential equation \eqref{SDE2} is seen as
\begin{equation}\label{SDE3}
d(f\circ q_t\circ h)=f_\ast V_0(f\circ q_t\circ h)dt+\sum _{k=1}^d f_\ast V_k(f\circ q_t\circ h)\circ dW^k_t,\hspace{0.3cm}f\circ q_0\circ h=\bar{q}
\end{equation}
on $M^{m+n+1}_{h_1,h_2}$. The properties of the flow $f\circ q_t\circ h$ can be understood from the equation \eqref{SDE3}, especially from the vector fields appearing in the equation.
At point $(\bar{x},\bar{y})\in M^{m+n+1}_{h_1,h_2}$ the value of vector field $f_\ast V_k$ is given by
\begin{equation}
(f_\ast V_k)(\bar{x}, \bar{y})=(f_\ast)_{(\gamma ,\kappa )}[V_k(\gamma ,\kappa )],
\end{equation}
where $(\gamma ,\kappa )=h(\bar{x},\bar{y})$. If $V_k^\mu $ $(\mu=1,2, \cdots , m+n+2)$ is the components of $V_k$ in the ambient space $\mathbb{R}^{m+n+1}$, then for $\nu =1,2,\cdots ,m+n+2$
\begin{equation}
{(f_\ast)_{(\gamma ,\kappa )}[V_k(\gamma ,\kappa )]}^\nu = \sum _{j=1}^{m+1} \frac{\partial f^\nu }{\partial \gamma ^j}V_k^j(\gamma ,\kappa )+\sum _{r=1}^{n+1} \frac{\partial f^\nu }{\partial \kappa ^r}V_k^r(\gamma ,\kappa ).
\end{equation}
If $(\bar{x},\bar{y})$ is in the region $S^m\times \mathbb{R}^{n+1}$ of $M^{m+n+1}_{h_1,h_2}$, the differentials of $f$ are given by
\begin{eqnarray}
\frac{\partial f^\nu}{\partial \gamma^j} & = &\sum_{i=1}^{m+1}\frac{\partial\:h_1^\nu(\hat{\gamma})}{\partial\hat{\gamma}^i}\;\left(\frac{\delta_j^i}{\|\gamma\|}-\frac{\|\kappa\|^2}{\|\gamma\|^3}\;{\gamma^i\gamma^j}\right)\nonumber \\
&=& \frac{\partial h_1^\nu }{\partial \hat{\gamma }^j}\sqrt{1+\|\bar{y}\|^2}-\sum _{i=1}^{m+1}\frac{\partial h_1^\nu }{\partial \hat{\gamma }^i}\frac{\|\bar{y}\|^2}{\sqrt{1+\|\bar{y}\|^2}}[h_1^{-1}(\bar{x})]^i[h_1^{-1}(\bar{x})]^j\qquad (\nu, j=1,..., m+1),\\
\frac{\partial f^\nu}{\partial \kappa^r}& = &\sum_{i=1}^{m+1}\frac{\partial\:h_1^\nu(\hat{\gamma})}{\partial\hat{\gamma}^i}\;\frac{\gamma^i\kappa^r}{\|\gamma\|}\nonumber \\
&=& \sum _{i=1}^{m+1}\frac{\partial h_1^\nu }{\partial \hat{\gamma }^i}\frac{\|\bar{y}\|}{\sqrt{1+\|\bar{y}\|^2}}[h_1^{-1}(\bar{x})]^i[h_2^{-1}(\bar{y}/\|\bar{y}\|)]^r\qquad (\nu=1,..., m+1,\; r=1,...,n+1),\\
\frac{\partial f^\nu}{\partial \gamma^j} &=&-\frac{\|\kappa\|}{\|\gamma\|^3}h_2^{\nu-m-1}((\hat{\kappa })\gamma^j+\sum_{s=1}^{n+1}\frac{\partial h_2^{\nu-m-1}(\hat{\kappa })}{\partial\hat{\kappa}^s}\frac{\kappa^s\gamma^j}{\|\gamma\|}\nonumber \\
&=&-\bar{y}^{\nu-m-1}\sqrt{1+\|\bar{y}\|^2}[h_1^{-1}(\bar{x})]^j+
\sum_{s=1}^{n+1}\frac{\partial h_2^{\nu-m-1}}{\partial\hat{\kappa}^s}\frac{[h_1^{-1}(\bar{x})]^j[h_2^{-1}(\bar{y}/\|\bar{y}\|)]^s}{\sqrt{1+\|\bar{y}\|^2}},
\end{eqnarray}
($j=1,...,m+1$, $\nu=m+2,..., m+n+2$), and
\begin{eqnarray}
\frac{\partial f^\nu}{\partial \kappa^r}&=&h_2^{\nu-m-1}(\hat{\kappa })\frac{\kappa^r}{ \|\gamma\| \|\kappa\|}+\sum_{s=1}^{n+1}\frac{\partial h_2^{\nu-m-1}(\hat{\kappa })}{\partial \hat{\kappa}^s}\left(\frac{\delta^s_r}{\|\gamma\|}-\frac{\|\gamma\|}{\|\kappa\|^2}{\kappa^s\kappa^r}\right)\nonumber \\
&=& \bar{y}^{\nu-m-1}\sqrt{1+\|\bar{y}\|^2}[h_2^{-1}(\bar{y}/\|\bar{y}\|)]^r \nonumber \\ & & +\sum _{s=1}^{n+1}\frac{\partial h_2^{\nu-m-1}}{\partial \hat{\kappa}^s}\left(\delta ^s_r\sqrt{1+\|\bar{y}\|^2}-\frac{[h_2^{-1}(\bar{y}/\|\bar{y}\|)]^s[h_2^{-1}(\bar{y}/\|\bar{y}\|)]^r}{\sqrt{1+\|\bar{y}\|^2}}\right),
\end{eqnarray}
($r=1,...,n+1$, $\nu =m+2,...,m+n+2$), since \begin{equation}h_2^{\nu-m-1}(\hat{\kappa })=h_2^{\nu-m-1}[h_2^{-1}(\bar{y}/\|\bar{y}\|)]=\frac{\bar{y}^{\nu-m-1}}{\|\bar{y}\|}.
\end{equation}
If $(\bar{x},\bar{y})$ is in the region $\mathbb{R}^{m+1}\times S^n$ of $M^{m+n+1}_{h_1,h_2}$, the differentials of $f$ are given by
\begin{eqnarray}
\frac{\partial f^\nu }{\partial \gamma ^j}&=&\frac{\gamma ^\nu \gamma ^j}{\|\kappa \|}+\frac{\delta ^\nu_j}{\|\kappa \|},\nonumber \\
&=&\frac{\bar{x}^\nu \bar{x}^j}{\sqrt{1+\|\bar{x}\|^2}}+\delta ^\nu _j \sqrt{1+\|\bar{x}\|^2}\qquad (\nu ,j=1,..., m+1),\\
\frac{\partial f^\nu }{\partial \kappa ^r}&=&-\frac{\|\gamma \|^2\gamma ^\nu \kappa  ^r}{\|\kappa \|^3}, \nonumber \\
&=&-\frac{\|\bar{x}\|^2\bar{x}^\nu \bar{y}^r}{\sqrt{1+\|\bar{x}\|^2}}\qquad (\nu =1,...,m+1,\; r=1,...,n+1),\\
\frac{\partial f^\nu }{\partial \gamma ^j}&=&\frac{\kappa ^{\nu-m-1} \gamma ^j}{\|\kappa \|},\nonumber \\
&=& \frac{\bar{x}^j\bar{y}^{\nu-m-1}}{\sqrt{1+\|\bar{x}\|^2}}\hspace{0.4cm} (\nu =m+2,...,m+n+2,\; j=1,..., m+1),\\
\frac{\partial f^\nu }{\partial \kappa ^r}&=&\frac{\delta ^{\nu-m-1} _r}{\|\kappa \|}-\frac{\|\gamma \|^2\kappa ^{\nu-m-1} \kappa ^r}{\|\kappa \|^3} \nonumber \\
&=&\sqrt{1+\|\bar{x}\|^2}\delta _r^{\nu -m-1}-\frac{\|\bar{x}\|^2\bar{y}^r\bar{y}^{\nu -m-1}}{\sqrt{1+\|\bar{x}\|^2}},\qquad \qquad \qquad
\end{eqnarray}
($\nu =m+2,...,m+n+2,\; r=1,...,n+1$).
In the above expressions, $\hat{\gamma }:=\gamma \|\gamma \|^{-1}=(\gamma ^1,...,\gamma ^{m+1})\|\gamma \|^{-1}$ and $\hat{\kappa }:=\kappa \|\kappa \|^{-1}=(\kappa ^1,...,\kappa ^{n+1})\|\kappa \|^{-1}$. Then, the vector fields $(f_\ast V_k)$ ($k=0,1,...,d$) at $(\bar{x}, \bar{y})\in S^m\times \mathbb{R}^{n+1}$ are given by
\begin{eqnarray}\label{vecMA}
[(f_\ast V_k)(\bar{x}, \bar{y})]^\nu&=& \sum _{j=1}^{m+1}\left[\frac{\partial h_1^\nu }{\partial \hat{\gamma }^j}\sqrt{1+\|\bar{y}\|^2}-\sum _{i=1}^{m+1}\frac{\partial h_1^\nu }{\partial \hat{\gamma }^i}\frac{\|\bar{y}\|^2}{\sqrt{1+\|\bar{y}\|^2}}[h_1^{-1}(\bar{x})]^i[h_1^{-1}(\bar{x})]^j\right]V_k^j(h(\bar{x},\bar{y}))\nonumber \\
& & +\sum _{r=1}^{n+1}\left[\sum _{i=1}^{m+1}\frac{\partial h_1^\nu }{\partial \hat{\gamma }^i}\frac{\|\bar{y}\|}{\sqrt{1+\|\bar{y}\|^2}}[h_1^{-1}(\bar{x})]^i[h_2^{-1}(\bar{y}/\|\bar{y}\|)]^r\right]V_k^r(h(\bar{x},\bar{y})),
\end{eqnarray}
for $\nu =1, ..., m+1$, and
\begin{eqnarray}\label{vecMB}
[(f_\ast V_k)(\bar{x}, \bar{y})]^\nu &=&\sum _{j=1}^{m+1}\left[-\bar{y}^{\nu-m-1}\sqrt{1+\|\bar{y}\|^2}[h_1^{-1}(\bar{x})]^j\right]V_k^j(h(\bar{x},\bar{y})) \nonumber \\ 
&& + \sum _{j=1}^{m+1} \left[\sum_{s=1}^{n+1}\frac{\partial h_2^{\nu-m-1}}{\partial\hat{\kappa}^s}\frac{[h_1^{-1}(\bar{x})]^j[h_2^{-1}(\bar{y}/\|\bar{y}\|)]^s}{\sqrt{1+\|\bar{y}\|^2}}\right] V_k^j(h(\bar{x},\bar{y}))\nonumber \\ 
&&+\sum _{r=1}^{n+1}\left[\bar{y}^{\nu-m-1}\sqrt{1+\|\bar{y}\|^2}[h_2^{-1}(\bar{y}/\|\bar{y}\|)]^r\right]V_k^r(h(\bar{x},\bar{y}))\nonumber \\
&&+\sum _{r=1}^{n+1}\left[\sum _{s=1}^{n+1}\frac{\partial h_2^{\nu-m-1}}{\partial \hat{\kappa}^s}\left(\delta ^s_r\sqrt{1+\|\bar{y}\|^2}-\frac{[h_2^{-1}(\bar{y}/\|\bar{y}\|)]^s[h_2^{-1}(\bar{y}/\|\bar{y}\|)]^r}{\sqrt{1+\|\bar{y}\|^2}}\right)\right]V_k^r(h(\bar{x},\bar{y})),\nonumber\\
\end{eqnarray}
for $\nu=m+2, ..., m+n+2$. Whereas, at $(\bar{x}, \bar{y})\in \mathbb{R}^{m+1}\times S^n$ the value of vector fields $(f_\ast V_k)$ ($k=0,1,...,d$) are given by
\begin{eqnarray}\label{vecM1}
[(f_\ast V_k)(\bar{x}, \bar{y})]^\nu & = & 
\sum _{j=1}^{m+1}\left[\frac{\bar{x}^\nu \bar{x}^j}{\sqrt{1+\|\bar{x}\|^2}}+\delta ^\nu _j \sqrt{1+\|\bar{x}\|^2}\right]V_k^j(h(\bar{x},\bar{y}))\nonumber \\
& &+ \sum _{r=1}^{n+1}\left[-\frac{\|\bar{x}\|^2\bar{x}^\nu \bar{y}^r}{\sqrt{1+\|\bar{x}\|^2}}\right]V_k^r(h(\bar{x},\bar{y})), \qquad \text{for }\nu = 1, ...,m+1
\end{eqnarray}

\begin{eqnarray}\label{vecM2}
[(f_\ast V_k)(\bar{x}, \bar{y})]^\nu &=&\sum _{j=1}^{m+1}\left[\frac{\bar{x}^j\bar{y}^{\nu-m-1}}{\sqrt{1+\|\bar{x}\|^2}}\right]V_k^j(h(\bar{x},\bar{y}))\nonumber \\
& & +\sum _{r=1}^{n+1}\left[\sqrt{1+\|\bar{x}\|^2}\delta _r^{\nu -m-1}-\frac{\|\bar{x}\|^2\bar{y}^r\bar{y}^{\nu -m-1}}{\sqrt{1+\|\bar{x}\|^2}}\right]V_k^r(h(\bar{x},\bar{y})),
\end{eqnarray}
for $\nu =m+2, ..., m+n+2$.

$\qquad$From the expressions of the components $[(f_\ast V_k)(\bar{x}, \bar{y})]^\nu $ of the vector fields $f_\ast V_k$ on the region $\mathbb{R}^{m+1}\times S^n$, i.e. Equation \eqref{vecM1} and \eqref{vecM2}, it is clear that the vector fields $f_\ast V_k$ ($k=1,...,d$) are $C^2$-differentiable on that region. 

$\qquad$Now we consider the expression of the components $[(f_\ast V_k)(\bar{x}, \bar{y})]^\nu $ (i.e. Equation \eqref{vecMA} and \eqref{vecMB}) of the vector fields $f_\ast V_k$ on the other region of $M^{m+n+1}_{h_1,h_2}$. From Equation \eqref{vecMA} and \eqref{vecMB}), the differentiability of the components $[(f_\ast V_k)(\bar{x}, \bar{y})]^\nu $ clearly depend on the differentiability of the komponents $V_k^\nu \circ h$ as well as on the differentiability of their koefisiens appearing in Equation \eqref{vecMA} and \eqref{vecMB}). Taking second partial derivative of $V_k^\nu \circ h$ with respect to $\bar{x}^j$ and $\bar{x}^l$ yields
\begin{eqnarray}
\frac{\partial ^2V_k^\nu (h(\bar{x},\bar{y}))}{\partial \bar{x}^j\partial \bar{x}^l}
& = & \sum _{i,a=1}^{m+1}\frac{\partial ^2V_k^\nu }{\partial \gamma ^a\partial \gamma ^i}\frac{\partial \gamma ^i}{\partial \bar{x}^j} \frac{\partial \gamma ^a}{\partial \bar{x}^l}
+\sum _{i=1}^{m+1}\frac{\partial V_k^\nu }{\partial \gamma ^i}\frac{\partial ^2\gamma ^i}{\partial \bar{x}^j\partial \bar{x}^l} \nonumber \\
& = &  \frac{1}{\|\bar{y}\|^2+1}\sum _{i,a=1}^{m+1}\frac{\partial ^2V_k^\nu }{\partial \gamma ^a\partial \gamma ^i}\frac{\partial [h_1^{-1}(\bar{x})]^a}{\partial \bar{x}^l}\frac{\partial [h_1^{-1}(\bar{x})]^i}{\partial \bar{x}^j}+\frac{1}{\sqrt{\|\bar{y}\|^2+1}}\sum _{i=1}^{m+1}\frac{\partial V_k^\nu }{\partial \gamma ^i}\frac{\partial ^2[h_1^{-1}(\bar{x})]^i}{\partial \bar{x}^j\partial \bar{x}^l}.\nonumber\\
\end{eqnarray}
It is clear that the differentiability of $V_k^\nu \circ h$ with respect to $\bar{x}$ depends only on the differentiability of $V_k^\nu $ with respect to $\gamma $, since $h_1$ is a diffeomorphism. Now taking second partial derivative of $V_k^\nu \circ h$ with respect to $\bar{y}^j$ and $\bar{x}^l$ yields
\begin{eqnarray}
\frac{\partial ^2V_k^\nu (h(\bar{x},\bar{y}))}{\partial \bar{y}^c\partial \bar{x}^l}
& = & \sum _{i,a=1}^{m+1}\frac{\partial ^2V_k^\nu }{\partial \gamma ^a\partial \gamma ^i}\frac{\partial \gamma ^i}{\partial \bar{y}^c} \frac{\partial \gamma ^a}{\partial \bar{x}^l}+\sum _{i=1}^{m+1}\frac{\partial V_k^\nu }{\partial \gamma ^i}\frac{\partial ^2\gamma ^i}{\partial \bar{y}^c\partial \bar{x}^l}+\sum _{a=1}^{m+1}\sum _{r=1}^{n+1}\frac{\partial ^2V_k^\nu }{\partial \gamma ^a\partial \kappa ^r}\frac{\partial \kappa ^r}{\partial \bar{y}^c} \frac{\partial \gamma ^a}{\partial \bar{x}^l}\nonumber \\
& = &  -\frac{1}{(\|\bar{y}\|^2+1)^2}\sum _{i,a=1}^{m+1}\frac{\partial ^2V_k^\nu }{\partial \gamma ^a\partial \gamma ^i}\frac{\partial [h_1^{-1}(\bar{x})]^a}{\partial \bar{x}^l}[h_1^{-1}(\bar{x})]^i\bar{y}^c\nonumber\\
& & -\frac{1}{(\|\bar{y}\|^2+1)^{3/2}}\sum _{i=1}^{m+1}\frac{\partial V_k^\nu }{\partial \gamma ^i}\frac{\partial [h_1^{-1}(\bar{x})]^i}{\partial \bar{x}^l}\bar{y}^c\nonumber\\
& &-\frac{1}{(\|\bar{y}\|^2+1)^2}\sum _{a=1}^{m+1}\sum _{r=1}^{n+1}\frac{\partial ^2V_k^\nu }{\partial \gamma ^a\partial \kappa ^r}\|\bar{y}\|[h_2^{-1}(\hat{\bar{y}})]^r \frac{\partial [h_1^{-1}(\bar{x})]^a}{\partial \bar{x}^l}\nonumber \\
& &+\frac{1}{\|\bar{y}\|^2+1}\sum _{a=1}^{m+1}\sum _{r=1}^{n+1}\frac{\partial ^2V_k^\nu }{\partial \gamma ^a\partial \kappa ^r}\frac{\bar{y}^c}{\|\bar{y}\|}[h_2^{-1}(\hat{\bar{y}})]^r \frac{\partial [h_1^{-1}(\bar{x})]^a}{\partial \bar{x}^l}\nonumber \\
& &-\frac{1}{\|\bar{y}\|^2+1}\sum _{a=1}^{m+1}\sum _{r=1}^{n+1}\frac{\partial ^2V_k^\nu }{\partial \gamma ^a\partial \kappa ^r}\sum _{s=1}^{n+1}\frac{\partial [h_2^{-1}(\hat{\bar{y}})]^r }{\partial \hat{\bar{y}}^s}\frac{\bar{y}^s\bar{y}^c}{\|\bar{y}\|^2} \frac{\partial [h_1^{-1}(\bar{x})]^a}{\partial \bar{x}^l}\nonumber \\
& &+\frac{1}{\|\bar{y}\|^2+1}\sum _{a=1}^{m+1}\sum _{r=1}^{n+1}\frac{\partial ^2V_k^\nu }{\partial \gamma ^a\partial \kappa ^r}\frac{\partial [h_2^{-1}(\hat{\bar{y}})]^r}{\partial \hat{\bar{y}}^c} \frac{\partial [h_1^{-1}(\bar{x})]^a}{\partial \bar{x}^l}.
\end{eqnarray}
It is also clear from the last equation that the second partial derivative of $V_k^\nu \circ h$ $(\nu )$ with respect to $\bar{x}$ and $\bar{y}$ is continuous only if the function $V_k^\nu $ is at least $C^2$. The second derivative of $V_k^\nu \circ h$ with respect to $\bar{y}$ is given by

\begin{eqnarray}
\frac{\partial ^2V_k^\nu (h(\bar{x},\bar{y}))}{\partial \bar{y}^c\partial \bar{y}^d}
& = & \sum _{i,j=1}^{m+1}\frac{\partial ^2V_k^\nu }{\partial \gamma ^j\partial \gamma ^i}\frac{\partial \gamma ^i}{\partial \bar{y}^c} \frac{\partial \gamma ^j}{\partial \bar{y}^d}+\sum _{i=1}^{m+1}\frac{\partial V_k^\nu }{\partial \gamma ^i}\frac{\partial ^2\gamma ^i}{\partial \bar{y}^c\partial \bar{y}^d}+\sum _{j=1}^{m+1}\sum _{r=1}^{n+1}\frac{\partial ^2V_k^\nu }{\partial \gamma ^j\partial \kappa ^r}\frac{\partial \kappa ^r}{\partial \bar{y}^c} \frac{\partial \gamma ^j}{\partial \bar{y}^d}\nonumber \\
& & +\sum _{s,r=1}^{n+1}\frac{\partial ^2V_k^\nu }{\partial \kappa ^r\partial \kappa ^s}\frac{\partial \kappa ^r}{\partial \bar{y}^c} \frac{\partial \kappa ^s}{\partial \bar{y}^d}+\sum _{j=1}^{m+1}\sum _{r=1}^{n+1}\frac{\partial ^2V_k^\nu }{\partial \gamma ^j\partial \kappa ^r}\frac{\partial \kappa ^r}{\partial \bar{y}^d} \frac{\partial \gamma ^j}{\partial \bar{y}^c}+\sum _{i=1}^{m+1}\frac{\partial V_k^\nu }{\partial \kappa ^r}\frac{\partial ^2\kappa ^r}{\partial \bar{y}^c\partial \bar{y}^d},
\end{eqnarray}
where
\begin{eqnarray}\label{koe1}
\frac{\partial \gamma ^i}{\partial \bar{y}^c} \frac{\partial \gamma ^j}{\partial \bar{y}^d}& = & \frac{[h_1^{-1}(\bar{x})]^i[h_1^{-1}(\bar{x})]^j\bar{y}^c\bar{y}^d}{(\|\bar{y}\|^2+1)^3}, 
\end{eqnarray}
\begin{eqnarray}\label{koe2}
\frac{\partial ^2\gamma ^i}{\partial \bar{y}^c\partial \bar{y}^d}
& = &
\frac{3[h_1^{-1}(\bar{x})]^i\bar{y}^c\bar{y}^d}{(\|\bar{y}\|^2+1)^{5/2}} -\frac{[h_1^{-1}(\bar{x})]^i\delta ^d_c}{(\|\bar{y}\|^2+1)^{3/2}},
\end{eqnarray}
\begin{eqnarray}\label{koe3}
\frac{\partial \kappa ^r}{\partial \bar{y}^c} \frac{\partial \gamma ^j}{\partial \bar{y}^d}
& = &  -\frac{\|\bar{y}\|[h_1^{-1}(\bar{x})]^j[h_2^{-1}(\hat{\bar{y}})]^r\bar{y}^c\bar{y}^d}{(\|\bar{y}\|^2+1)^3} +\frac{[h_1^{-1}(\bar{x})]^j[h_2^{-1}(\hat{\bar{y}})]^r}{(\|\bar{y}\|^2+1)^2}\frac{\bar{y}^c\bar{y}^d}{\|\bar{y}\|}\nonumber \\
& & -\sum_{s=1}^{n+1}\frac{[h_1^{-1}(\bar{x})]^j}{(\|\bar{y}\|^2+1)^2}\frac{\partial [h_2^{-1}(\hat{\bar{y}})]^r }{\partial \hat{\bar{y}}^s}\frac{\bar{y}^d\bar{y}^s\bar{y}^c}{\|\bar{y}\|^2} +\frac{[h_1^{-1}(\bar{x})]^j\bar{y}^d}{(\|\bar{y}\|^2+1)^2}\frac{\partial [h_2^{-1}(\hat{\bar{y}})]^r}{\partial \hat{\bar{y}}^c},
\end{eqnarray}
\begin{eqnarray}\label{koe4}
\frac{\partial \kappa ^r}{\partial \bar{y}^d} \frac{\partial \gamma ^j}{\partial \bar{y}^c}
& = &  -\frac{\|\bar{y}\|[h_1^{-1}(\bar{x})]^j[h_2^{-1}(\hat{\bar{y}})]^r\bar{y}^c\bar{y}^d}{(\|\bar{y}\|^2+1)^3} +\frac{[h_1^{-1}(\bar{x})]^j[h_2^{-1}(\hat{\bar{y}})]^r}{(\|\bar{y}\|^2+1)^2}\frac{\bar{y}^c\bar{y}^d}{\|\bar{y}\|}\nonumber \\
& & -\sum_{s=1}^{n+1}\frac{[h_1^{-1}(\bar{x})]^j}{(\|\bar{y}\|^2+1)^2}\frac{\partial [h_2^{-1}(\hat{\bar{y}})]^r }{\partial \hat{\bar{y}}^s}\frac{\bar{y}^d\bar{y}^s\bar{y}^c}{\|\bar{y}\|^2} +\frac{[h_1^{-1}(\bar{x})]^j\bar{y}^c}{(\|\bar{y}\|^2+1)^2}\frac{\partial [h_2^{-1}(\hat{\bar{y}})]^r}{\partial \hat{\bar{y}}^d},
\end{eqnarray}

\begin{eqnarray}\label{koe5}
\frac{\partial \kappa ^r}{\partial \bar{y}^c} \frac{\partial \kappa ^s}{\partial \bar{y}^d} &=& \frac{\|\bar{y}\|^2[h_2^{-1}(\hat{\bar{y}})]^r[h_2^{-1}(\hat{\bar{y}})]^s\bar{y}^c\bar{y}^d}{(\|\bar{y}\|^2+1)^3}
+\frac{[h_2^{-1}(\hat{\bar{y}})]^r[h_2^{-1}(\hat{\bar{y}})]^s\bar{y}^c\bar{y}^d}{(\|\bar{y}\|^2+1)\|\bar{y}\|^2}\nonumber \\
& &
+\frac{1}{\|\bar{y}\|^2+1}\frac{\partial[h_2^{-1}(\hat{\bar{y}})]^r }{\partial \hat{\bar{y}}^c}\frac{\partial[h_2^{-1}(\hat{\bar{y}})]^s }{\partial \hat{\bar{y}}^d}\nonumber \\
& &
+\frac{1}{\|\bar{y}\|^2+1}\sum _{t,u=1}^{n+1}\frac{\partial[h_2^{-1}(\hat{\bar{y}})]^r }{\partial \hat{\bar{y}}^t}\frac{\partial[h_2^{-1}(\hat{\bar{y}})]^s }{\partial \hat{\bar{y}}^u}\frac{\bar{y}^u\bar{y}^t\bar{y}^c\bar{y}^d}{\|\bar{y}\|^4}\nonumber \\
& &
-2\frac{[h_2^{-1}(\hat{\bar{y}})]^r[h_2^{-1}(\hat{\bar{y}})]^s\bar{y}^c\bar{y}^d}{(\|\bar{y}\|^2+1)^2}
-\frac{\|\bar{y}\|[h_2^{-1}(\hat{\bar{y}})]^r\bar{y}^c}{(\|\bar{y}\|^2+1)^2}\frac{\partial[h_2^{-1}(\hat{\bar{y}})]^s }{\partial \hat{\bar{y}}^d}\nonumber \\
& &
+\frac{[h_2^{-1}(\hat{\bar{y}})]^r\bar{y}^c\bar{y}^d}{(\|\bar{y}\|^2+1)^2\|\bar{y}\|}\sum _{t=1}^{n+1}\frac{\partial[h_2^{-1}(\hat{\bar{y}})]^s }{\partial \hat{\bar{y}}^t}\bar{y}^t
+\frac{[h_2^{-1}(\hat{\bar{y}})]^r\bar{y}^c}{(\|\bar{y}\|^2+1)\|\bar{y}\|}\frac{\partial[h_2^{-1}(\hat{\bar{y}})]^s }{\partial \hat{\bar{y}}^d}\nonumber \\
& &
-\frac{[h_2^{-1}(\hat{\bar{y}})]^r}{\|\bar{y}\|^2+1}\sum _{t=1}^{n+1}\frac{\partial[h_2^{-1}(\hat{\bar{y}})]^s }{\partial \hat{\bar{y}}^t}\frac{\bar{y}^t\bar{y}^c\bar{y}^d}{\|\bar{y}\|^3}
+\frac{[h_2^{-1}(\hat{\bar{y}})]^s\bar{y}^d}{(\|\bar{y}\|^2+1)^2\|\bar{y}\|}\frac{\partial[h_2^{-1}(\hat{\bar{y}})]^r }{\partial \hat{\bar{y}}^c}\nonumber \\
& &
-\frac{1}{\|\bar{y}\|^2+1}\frac{\partial [h_2^{-1}(\hat{\bar{y}})]^r}{\partial \hat{\bar{y}}^c}\sum _{t=1}^{n+1}\frac{\partial[h_2^{-1}(\hat{\bar{y}})]^s }{\partial \hat{\bar{y}}^t}\frac{\bar{y}^t\bar{y}^d}{\|\bar{y}\|^2}\nonumber \\
& &
-\frac{1}{\|\bar{y}\|^2+1}\frac{\partial [h_2^{-1}(\hat{\bar{y}})]^s}{\partial \hat{\bar{y}}^d}\sum _{t=1}^{n+1}\frac{\partial[h_2^{-1}(\hat{\bar{y}})]^r }{\partial \hat{\bar{y}}^t}\frac{\bar{y}^t\bar{y}^c}{\|\bar{y}\|^2}\nonumber \\
& &
+\frac{[h_2^{-1}(\hat{\bar{y}})]^s}{(\|\bar{y}\|^2+1)^2}\sum _{t=1}^{n+1}\frac{\partial[h_2^{-1}(\hat{\bar{y}})]^r }{\partial \hat{\bar{y}}^t}\frac{\bar{y}^t\bar{y}^c\bar{y}^d}{\|\bar{y}\|^3},
\end{eqnarray}
and
\begin{eqnarray}
\frac{\partial ^2\kappa ^r}{\partial \bar{y}^c\partial \bar{y}^d}&=&
\frac{3\|\bar{y}\|[h_2^{-1}(\hat{\bar{y}})]^r\bar{y}^c\bar{y}^d}{(\|\bar{y}\|^2+1)^{5/2}}
-\frac{\bar{y}^c}{(\|\bar{y}\|^2+1)^{3/2}}\frac{\partial [h_2^{-1}(\hat{\bar{y}})]^r}{\partial \hat{\bar{y}}^d}\nonumber \\
& &
+\frac{1}{(\|\bar{y}\|^2+1)^{3/2}}\sum _{s=1}^{n+1}\frac{\partial [h_2^{-1}(\hat{\bar{y}})]^r}{\partial \hat{\bar{y}}^s}\frac{\bar{y}^c\bar{y}^s\bar{y}^d}{\|\bar{y}\|^2}
-\frac{\|\bar{y}\|[h_2^{-1}(\hat{\bar{y}})]^r\delta ^c_d}{(\|\bar{y}\|^2+1)^{3/2}}\nonumber \\
& & -\frac{[h_2^{-1}(\hat{\bar{y}})]^r\bar{y}^c\bar{y}^d}{(\|\bar{y}\|^2+1)^{3/2}\|\bar{y}\|}
-\frac{[h_2^{-1}(\hat{\bar{y}})]^r\bar{y}^c\bar{y}^d}{(\|\bar{y}\|^2+1)^{3/2}\|\bar{y}\|}
+\frac{[h_2^{-1}(\hat{\bar{y}})]^r\delta ^c_d}{\|\bar{y}\|\sqrt{\|\bar{y}\|^2+1}}\nonumber \\
& &
+\frac{1}{\sqrt{\|\bar{y}\|^2+1}}\frac{\partial [h_2^{-1}(\hat{\bar{y}})]^r}{\partial \hat{\bar{y}}^d}\frac{\bar{y}^c}{\|\bar{y}\|^2}
-\frac{1}{\sqrt{\|\bar{y}\|^2+1}}\sum _{s=1}^{n+1}\frac{\partial [h_2^{-1}(\hat{\bar{y}})]^r}{\partial \hat{\bar{y}}^s}\frac{\bar{y}^c\bar{y}^s\bar{y}^d}{\|\bar{y}\|^4}
\nonumber \\
& & 
-\frac{[h_2^{-1}(\hat{\bar{y}})]^r\bar{y}^c\bar{y}^d}{\|\bar{y}\|^3\sqrt{\|\bar{y}\|^2+1}}
-\frac{1}{(\|\bar{y}\|^2+1)^{3/2}}\frac{\partial [h_2^{-1}(\hat{\bar{y}})]^r}{\partial \hat{\bar{y}}^c}\frac{\bar{y}^d}{\|\bar{y}\|}\nonumber \\
& &
+\frac{1}{(\|\bar{y}\|^2+1)^{3/2}}\sum _{s=1}^{n+1}\frac{\partial [h_2^{-1}(\hat{\bar{y}})]^r}{\partial \hat{\bar{y}}^s}\frac{\bar{y}^s\bar{y}^c\bar{y}^d}{\|\bar{y}\|^3}\nonumber\\
\label{koe6}
& &
+\frac{1}{\sqrt{\|\bar{y}\|^2+1}}\frac{\partial ^2[h_2^{-1}(\hat{\bar{y}})]^r}{\partial \hat{\bar{y}}^d\partial \hat{\bar{y}}^c}\frac{1}{\|\bar{y}\|}
-\frac{1}{\sqrt{\|\bar{y}\|^2+1}}\sum _{s=1}^{n+1}\frac{\partial ^2[h_2^{-1}(\hat{\bar{y}})]^r}{\partial \hat{\bar{y}}^s\partial \hat{\bar{y}}^c}\frac{\bar{y}^s\bar{y}^d}{\|\bar{y}\|^3}\nonumber \\
& &
+\frac{1}{(\|\bar{y}\|^2+1)^{3/2}}\sum _{s=1}^{n+1}\frac{\partial [h_2^{-1}(\hat{\bar{y}})]^r}{\partial \hat{\bar{y}}^s}\frac{\bar{y}^s\bar{y}^c\bar{y}^d}{\|\bar{y}\|^2}\nonumber\\
& &
+\frac{2}{\sqrt{\|\bar{y}\|^2+1)}}\sum _{s=1}^{n+1}\frac{\partial [h_2^{-1}(\hat{\bar{y}})]^r}{\partial \hat{\bar{y}}^s}\frac{\bar{y}^s\bar{y}^c\bar{y}^d}{\|\bar{y}\|^4}\nonumber \\
& & 
-\frac{\delta ^c_d}{\sqrt{\|\bar{y}\|^2+1)}}\sum _{s=1}^{n+1}\frac{\partial [h_2^{-1}(\hat{\bar{y}})]^r}{\partial \hat{\bar{y}}^s}\frac{\bar{y}^s}{\|\bar{y}\|^2}
-\frac{1}{\sqrt{\|\bar{y}\|^2+1)}}\frac{\partial [h_2^{-1}(\hat{\bar{y}})]^r}{\partial \hat{\bar{y}}^d}\frac{\bar{y}^c}{\|\bar{y}\|^2}\nonumber \\
& &-\frac{1}{\sqrt{\|\bar{y}\|^2+1}}\sum _{s=1}^{n+1}\frac{\partial ^2[h_2^{-1}(\hat{\bar{y}})]^r}{\partial \hat{\bar{y}}^d\partial \hat{\bar{y}}^s}\frac{\bar{y}^s\bar{y}^c}{\|\bar{y}\|^3}\nonumber \\
& &
-\frac{1}{\sqrt{\|\bar{y}\|^2+1}}\sum _{s,t=1}^{n+1}\frac{\partial ^2[h_2^{-1}(\hat{\bar{y}})]^r}{\partial \hat{\bar{y}}^t\partial \hat{\bar{y}}^s}\frac{\bar{y}^t\bar{y}^s\bar{y}^c\bar{y}^d}{\|\bar{y}\|^5}.
\end{eqnarray}

$\qquad$All expresions appearing in Equation \eqref{koe1}, \eqref{koe2}, \eqref{koe3}, \eqref{koe4}, and \eqref{koe5} are continuous. However, the continuity of the expresion appearing in Equation \eqref{koe6} very depends on the form of the diffeomorphism $h_2$. By calculating the second derivative of $\|\bar{y}\|[h_2^{-1}(\bar{y}/\|\bar{y}\|)]^r$ with respect to $\bar{y}^c$ and $\bar{y}^d$ ($c,d=1,2\cdots, n+1$) it is clear that if $\|\bar{y}\|[h_2^{-1}(\bar{y}/\|\bar{y}\|)]^r$ is $C^2$-differentiable with respect to $\bar{y}$ on the region $S^m\times \mathbb{R}^{n+1}$, then the expression appearing in Equation \eqref{koe6} continuous on the region and therefore all functions $V_k^\nu \circ h$ are $C^2$-differentiable. By inspecting the coefisiens of the functions $V_k^\nu \circ h$ in Equation \eqref{vecMA}, it easy to see that the components $[(f_\ast V_k)]^\nu$ ($\nu =1,2, \cdots ,m+1$) of the vector fields $[(f_\ast V_k)]$ are $C^2$-differentiable on $S^m\times \mathbb{R}^{n+1}$ if $\|\bar{y}\|[h_2^{-1}(\bar{y}/\|\bar{y}\|)]^r$ ($r=1,2, \cdots n+1$) are $C^2$-differentiable with respect to $\bar{y}$ on the region. By direct calculation, it can be concluded that if the factors $[h_2^{-1}(\bar{y}/\|\bar{y}\|)]^r$ are $C^2$-differentiable with respect to $\bar{y}$ for all $r=1,\cdots ,n+1$, then $\|\bar{y}\|[h_2^{-1}(\bar{y}/\|\bar{y}\|)]^r$ and $\bar{y}^s[h_2^{-1}(\bar{y}/\|\bar{y}\|)]^r$, for all $r,s=1,\cdots ,n+1$, are also $C^2$-differentiable with respect to $\bar{y}$. Furthermore, by inspecting the coefisiens of $V_k^\nu \circ h$ in Equation \eqref{vecMB}, the components $[(f_\ast V_k)]^\nu$ ($\nu =m+2, \cdots ,m+n+2$) of the vector fields $[(f_\ast V_k)]$ are $C^2$-differentiable on the region $S^m\times \mathbb{R}^{n+1}$ if the factors $[h_2^{-1}(\bar{y}/\|\bar{y}\|)]^r$ are $C^2$-differentiable on the region for all $r=1,2, \cdots n+1$.

$\qquad$Now, we can extend the map $f:S^{m+n+1}_s\rightarrow M^{m+n+1}_{h_1,h_2}$ defined in Equation \eqref{definisif} to a homeomorphism $f_{ex}$ from an open subset $U$ containing $S^{m+n+1}_s$ of $\mathbb{R}^{m+n+2}$ onto an open subsets $U'$ containing $M^{m+n+1}_{h_1,h_2}$ of $\mathbb{R}^{m+n+2}$. Let $\iota_S$ be the natural inclusion of $S^{m+n+1}_s$ into $\mathbb{R}^{m+n+2}$ and $G$ a Riemannian metric tensor in the ambient space $\mathbb{R}^{m+n+2}$. Furthermore, let $g$ denote the Riemannian metric tensor on $S^{m+n+1}_s$ inherited from $G$, i.e. $g=\iota _S^\ast G$. By using the map $h_{ex}$, i.e. the invers of $f_{ex}$, we can pull back the metric tensor field $G$ yielding the Riemannian metric tensor $h_{ex}^{\ast }G$ in $U'$. Let $\iota _M$ be the natural inclusion of $M^{m+n+1}_{h_1,h_2}$ into $\mathbb{R}^{m+n+2}$. Then, the pull back $\iota _M^\ast h_{ex}^{\ast }G$ of $h_{ex}^{\ast }G$ by $\iota _M$ is a Riemannian tensor metric on $M^{m+n+1}_{h_1,h_2}$. It is easy to show that the metric tensor $\iota _M^\ast h_{ex}^{\ast }G$ is the metric tensor $h^\ast g=h^\ast (\iota _S^\ast G)$. The metric tensor $\iota _M^\ast h_{ex}^{\ast }G$ can be regarded as the metric tensor $g$ on $S^{m+n+1}$ when it is described with the exotic differential structure. The metric tensor field $g$ gives rise to the Riemannian measures $\nu $ on $S^{m+n+1}_s$ and the metric tensor field $h^\ast g$ to the Riemannian measure $h^\ast \nu$ on $M^{m+n+1}_{h_1,h_2}$ respectivelly. 
Now define
\begin{equation}
|f_\ast X((\bar{q}))|_{h^\ast g}:=(h^\ast g)(\bar{q})((f_\ast X)(\bar{q}),(f_\ast X)(\bar{q})),
\end{equation}
for every vector field $X$ defined on $M^{m+n+1}_{h_1,h_2}$. It is straightforward to show that
\begin{equation}
|f_\ast X((\bar{q}))|_{h^\ast g}=|X|_g(h(\bar{q})).
\end{equation}
Since all the elements of the matrix representing the pull-back map $h^\ast$ as well as of the push-forward map $f_\ast $ are continuous then
\begin{equation}
\int _{M^{m+n+1}_{h_1,h_2}}|f_\ast V_0((\bar{q}))|^p(h^\ast \nu ) (d\bar{q})<\infty,
\end{equation}
whenever
\begin{equation}
\int _{S^{m+n+1}_s}|V_0(q)|^p\nu (dq)<\infty,
\end{equation}
for $1\geq p$, i.e. $f_\ast V_0\in L^p(TM^{m+n+1}_{h_1,h_2})$.

$\qquad$Since $V_0\in \mathbb{H}^p_1(TS^{m+n+1}_s)$, then there exist $m+n+1$ vector fields $Y_1,...,Y_{m+n+1}$ on $S^{m+n+1}_s$ so that for $\sigma = 1,\cdots , m+n+1$ we have
\begin{equation}\label{weakderivative}
\int _{S^{m+n+1}_s}\frac{\partial \varphi }{\partial q^\sigma }V_0(q)\nu (dq)=-\int _{S^{m+n+1}_s}\varphi (q)Y_\sigma (q)\nu (dq),
\end{equation}
(the integrations appearing in Equation \eqref{weakderivative} are in the Bochner sense) for every $\varphi \in C_c^\infty (S^{m+n+1}_s)$ and
\begin{equation}
\int _{S^{m+n+1}_s}|\nabla V_0 (q)|_g^p\nu (dq)<\infty ,
\end{equation}
where
\begin{equation}\label{sobolev}
|\nabla V_0(q)|_g=\left(\sum _{\eta =1}^{m+n+1}|Y_\eta (q)|_g^2\right)^{1/2}=\left(\sum _{\eta =1}^{m+n+1}g(q)(Y_\eta (q),Y_\eta (q)
)\right)^{1/2}.
\end{equation}
From Equation \eqref{sobolev} we obtain
\begin{equation}\label{sobolev1}
|f_\ast \nabla V_0(\bar{q})|_{h^\ast g}=\left(\sum _{\eta =1}^{m+n+1}(h^\ast g)(\bar{q})(f_\ast Y_\eta (\bar{q}),f_\ast Y_\eta (\bar{q})
)\right)^{1/2}=|\nabla V_0(\bar{q})|_g(h(\bar{q})).
\end{equation}
Since all the elements of the matrix representing the pull-back map $h^\ast$ are continuous, then
\begin{equation}
\int _{M^{m+n+1}_{h_1h_2}}|f_\ast \nabla V_0 (\bar{q})|^p(h^\ast \nu )(d\bar{q})<\infty,
\end{equation}
and therefore $f_\ast V_0$ is contained in the Sobolev space $\mathbb{H}^p_1(TM^{m+n+1}_{h_1,h_2})$.

$\qquad$Furthermore, if $V_0$ is in the space $L^\infty (TS_s^{m+n+1})$, then $f_\ast V_0$ belongs to the space $L^\infty (TM^{m+n+1}_{h_1,h_2})$ of all bounded measurable vector fields on $M^{m+n+1}_{h_1,h_2}$, since the map $f_\ast $ is continuous. Now because $S^{m+n+1}_s$ as well as $M^{m+n+1}_{h_1,h_2}$ is compact and $V_0\in L^\infty (TS_s^{m+n+1})$, it is clear that $V_0$ is in $ L^1 (TS_s^{m+n+1})$ and $f_\ast V_0$ is in $L^1(TM^{m+n+1}_{h_1,h_2})$. If $V_0$ has (weak) divergence with $\text{div }V_0\in L^\infty (S^{m+n+1}_s)$, then so is $f_\ast V_0$ with $\text{div }(f_\ast V_0)\in L^\infty (M^{m+n+1}_{h_1,h_2})$, since the maps $f_\ast $ and $h^\ast $ are continuous.

$\qquad$As a conclusion we have
\begin{theorem}\label{theorem2}
	Let $M^{m+n+1}_{h_1,h_2}$ be an exotic sphere, where $h_2^{-1}(\bar{y}/\|\bar{y}\|)$ is $C^2$-differentiable with respect to $\bar{y}$. Furthermore, let $V_0\in L^\infty (TS^{m+n+1}_s)\cap \mathbb{H}^p_1(TS^{m+n+1}_s)$ for some $p\geq 1$, $\text{div } V_0\in L^\infty (S^{m+n+1}_s)$, and for each $k=1,\cdots ,d$, $V_k\in C^2(TS^{m+n+1}_s)$ so that $q_t(q)$ is the unique $\nu$-almost everywhere stochastic invertible flow of \eqref{SDE2}. Then, the flow $(f\circ q_t\circ h)(\bar{q})$ on $M^{m+n+1}_{h_1,h_2}$ is the unique $h^\ast \nu$-almost everywhere stochastic invertible flow of \eqref{SDE3} in the sense of Definition \ref{DefZhang}
\end{theorem}

$\qquad$With the same conditions on the vector fields $V_k$ $(k=0,1,\cdots,d)$, Theorem \ref{theorem2} means that whenever $h_2^{-1}(\bar{y}/\|\bar{y}\|)$ is $C^2$-differentiable with respect to $\bar{y}$, both stochastic flows $q_t(q)$ and $(f\circ q_t\circ h)(\bar{q})$ have the same regularities. It also means, both description (using the standard diferential structure as well as using the choosen exotic differential structure) of the stochastic flow on the sphere $S^{m+n+1}$ look the same or indistinguishable. If both description of the stochastic flow on the sphere $S^{m+n+1}$ are distinguishable, then the diffeomorphism $h_2^{-1}(\bar{y}/\|\bar{y}\|)$ is not $C^2$-differentiable with respect to $\bar{y}$. Now if the map $h_2^{-1}(\bar{y}/\|\bar{y}\|)$ is {\em not} $C^2$-differentiable with respect to $\bar{y}$, then both description of the stochastic flow on the sphere $S^{m+n+1}$ may or may not indistinguishable.


\begin{thebibliography}{1}
\bibitem{Torsten}
Asselmeyer-Maluga, T., Brans, C.H.: {Exotic Smoothness and Physics: Differential Topology and Spacetime Models}. World Scientific, Singapore (2007)

\bibitem{Brans}
Brans, C.H.: Exotic Smoothness and physics. {J. Math. Phys.} {\textbf{35}}, 5494-5506 (1994)


\bibitem{Englert}
Englert, F., Rooman, M., Spindel, P.: Supersymmetry Breaking by Torsion and Ricci-flat Squeshed Seven-Sphere. Phys. Lett. {\textbf{B127}}, 47-50 (1983)


\bibitem{paper Freedman}
Freedman, M.: The topology of four dimensional manifolds. J. Differential Geom. {\bf 17}, 357-453 (1982)


\bibitem{book Hsu}
Hsu, E.P.: {Stochastic Analysis on Manifolds}. American Mathematical Society, Providence, Rhode Island (2002)



\bibitem{book Ikeda and watanabe}
Ikeda, N., Watanabe, S.: {Stochastic Differential Equations and Diffusion Processes}, Second edition, North-Holland Mathematical Library, {\bf 24}. North-Holland Publishing Company, North-Holland/ Kodansha (1989)


\bibitem{Kengo2}
Inami, T., Yamagishi, K.: Vanishing quantum vacuum energy in eleven-dimensional supergravity on the round seven-sphere. {Phys. Lett.} {\textbf{B143}}, 115-120 (1984)



\bibitem{paper Milnor}
Milnor, J.W.: On manifolds homeomorphic to the $7$-sphere. Ann. of Math. {\textbf{64}}, 399-405 (1956)


\bibitem{paper Milnor1}
Milnor, J.W.: Differentiable structures on spheres. {Amer. J. Math.} {\textbf{81}}, 962-972 (1959)



\bibitem{Edward Witten}
Witten, E.: Global gravitational anomalies, {Comm. Math. Phys.} {\textbf{100}}, 197-229 (1985)


\bibitem{Kengo1}
Yamagishi, K.: Supergravity on seven-dimensional homotopy spheres, { Phys. Lett. B}. {\bf 134},  47-50 (1984)


\bibitem{Zhang}
Zang, X.: Quasi-invariant stochastic flows of SDEs with non-smooth drifts on compact manifolds. {Stochastic Process. Appl.}{ \textbf{121}}, 1373-1388 (2011)
	
\end{thebibliography}
\end{document}